\documentclass[11pt,dvips]{article}
\usepackage{epsfig,times}
\usepackage{picinpar}
\makeatletter
\renewcommand{\section}{\@startsection{section}{1}{0in}
	{0.4\baselineskip}{0.1\baselineskip}{\Large\bf}}
\renewcommand{\subsection}{\@startsection{subsection}{2}{0in}
	{0.25\baselineskip}{-\baselineskip}{\large\bf}}
\renewcommand{\subsubsection}{\@startsection{subsubsection}{3}{0in}
	{0.1\baselineskip}{-\baselineskip}{\normalsize\bf}}
\makeatother

\begin{document}
\thispagestyle{myheadings}
\markright{OG 2.4.20}
\begin{center}
{\LARGE \bf Observations of the Crab Nebula, Mkn 501 and Mkn 421 using the TACTIC Imaging Element }
\end{center} 

\begin{center}

{\bf N. Bhatt, N.K. Agarwal, C.K. Bhat, S. Bhattacharyya, V.K. Dhar, A. Goyal, H.C. Goyal, C.L. Kaul, D.K. Koul, I.K. Kaul, R.K. Kaul,
S.K. Kaul, S.R. Kaul, M.K. Koul, R. Koul, M. Kothari, R.C. Rannot, A.K. Razdan,
S. Sahayanathan, M.L. Sapru, N. Satyabhama, A.K. Tickoo, N. Venugopal, K.K. Yadav
and C.L. Bhat}\\
{\it Bhabha Atomic Research Centre, Nuclear Research Laboratory, Mumbai-400085,
India}
\end{center}

\begin{center}
{\large \bf Abstract\\}
\end{center}
\vspace{-0.5ex}

The results of our observational campaigns on the two
extragalactic sources Mkn 501 and Mkn 421, carried out with the 
Imaging Element of the TACTIC array, during March-May, 1998 and April-May,
1999, are presented.  The results indicate that the two BL Lac objects
(Mkn 501 and Mkn 421) were in a `low' gamma-ray emission state during
both epochs of our observations.

\vspace{1ex}
\section{Introduction:}
\label{intro.sec}
The Imaging Element (IE) of the 4-element TACTIC array (Bhat, 1996) has seen
first light in March 1997, when it was used to observe the Crab Nebula and
the two BL Lac objects, Mkn 421 and Mkn 501, with a prototype 81-pixel
Cerenkov Light Imaging Camera.  The results of these observations,
including the successful detection of a TeV gamma-ray signal from the Mkn 501,
were reported at the 25th ICRC (Bhat et al, 1997; Protheroe et al, 1997).
The prototype camera is in the process of being upgraded to its full complement
of 349 pixels and is presently operational with a 169-pixel camera.  During
March - May, 1998, the IE was redeployed, whenever available, for
observations on the two BL Lac objects, with its 81-pixel camera,
as a part of the international multiwavelength observation campaign,
while for the 1999 observations the augmented 169-pixel camera was
used in order to improve the statistics.

As referred to in Koul et al, (1997), the IE employs 32
$\times$ 0.6 m spherical mirrors mounted on a Davies-Cotton structure,
giving an overall collection area of $\sim 9.5 m^2$.  During the 1998
observations, the central 9 $\times$ 9 pixels of the camera were
activated, leading to a field-of-view of $2.8^o \times 2.8^o$ with a pixel
resolution of $\sim 0.31^o$.  The camera size was increased to 169 pixels
(13 $\times$ 13) for the 1999 observations leading to a field-of-view of 
$\sim 4^o \times 4^o$.  The charge-to-digital (CDC) counts
of all the active pixels are logged following a master system trigger
which is derived from the innermost 36/144 pixels (for 1998/1999 observations)
by demanding a prompt coincidence (resolving time $\sim$ 10 ns) through
the 3 NCT proximity hardware trigger (Bhat et al, 1994).  The details
about observational methodology, system calibration and
data reduction and analysis technique are available in Bhat et al (1997).

\section{Observations and Data Treatment:}
During the 1998 observational spell, it became possible to use
the IE to observe the Crab
Nebula for only $\sim$ 10 h in the on-off mode of observations.
The resulting poor statistics and the low flux sensitivity of the 9 x 9 pixel interim camera, used for the
Crab observations does not allow a meaningful analysis of the
Crab data base and is, therefore, not discussed further.
The two extragalactic sources Mkn 421 (34.8 h) and
Mkn 501 ($\sim$ 52 h) were tracked upto source zenith angle $\theta \le 45^o$.  No
off-source scans were made in order to maximize the on-source observation time
and to improve the chances of recording possible flaring activity from these sources.  During the observational spell in March-May 1999,
the augmented TACTIC Imaging Element, with a 169 pixel camera, was again
deployed to observe the extragalactic
sources Mkn 421 (20 h) and Mkn 501 ( 16.2 h).  The daily data
stretches were subjected to the standard image-processing treatment in
which each raw image was first cleaned to minimize the noise component in
the event, using a graded noise-filter cut (Bhat et al, 1997).  The CDC
counts of the surviving `clean pixels' were flat-fielded by normalizing
their noise-subtracted CDC counts using the calibration data.  All images
with $\le$ 4 ' active ' pixels with non-zero flat-fielded counts or with $\ge$ 1
pixel with saturation-level CDC counts ($\ge$ 4090) were rejected.  For
the surviving events, the image parameters Length (L), Width (W),
Azwidth (A), Distance (D) and orientation parameter ($\alpha$) were
calculated and the total normalized count, (size S), determined
for each processed image.  Based on detailed Monte Carlo simulations
of the TACTIC system (Sapru et al, 1997), the following filters
were used for $\gamma$/h
separation: $0.10^o \le L \le 0.31^o$; $0.08^o \le W \le 0.17^o$; $0.55^o
\le D \le 1.10^o$; size $\ge$ 500.

\section{Results:}

%\subsection{1998 and 1999 data sets:}
The minimum image size value of $S_o$=500 CDC counts ($\sim$ 80-100
photoelectrons) was demanded to ensure a robust image quality and the
numbers of processed images surviving the above mentioned image parameter
cuts determined.  Table given below  gives a summary of the results obtained for Mkn
421 and Mkn 501, for the 1998 data set.\\
\begin{tabular}{|lllll|}\hline
   &Parameter&Mkn501(I)&Mkn501(II)&Mkn421\\
1. & Epoch of obsv. & April 23 & April 12 &  March 23 \\
- &-&- May 29, 98 & - April 22, 99 & - May 27, 98\\
2. & \# Observation days & 18 & 11 & 21\\
3. & Total Observation Time ( h )  & 3124.0 & 973.7 & 2089.2\\
4. & Total Processed images & 461796 & 254978 & 182963\\
5. & \#  Selected Events & 3820 & 877 & 2576\\
6. & \#  Events in $\gamma$ domain ($\alpha < 15^o$ ) & 686.00 $\pm $26.19 & 144.00 $\pm$ 12.00 & 486.0 $\pm$ 22.05\\
7. & \#  Background Events ($\alpha < 15^o$ )  & 645.00 $\pm$ 14.66 & 146.33 $\pm$ 6.98 & 437.33 $\pm$ 12.07\\
8. & \#  Excess Events & 41.00 $\pm$ 30.02 & -2.33 $\pm$ 13.88 & 48.67 $\pm$ 25.14\\
9. & Significance ($\sigma$) & 1.37 & -0.17 & 1.94\\ \hline
\end{tabular}\\
\begin{figwindow}[1,r,%
{\mbox{\epsfig{file=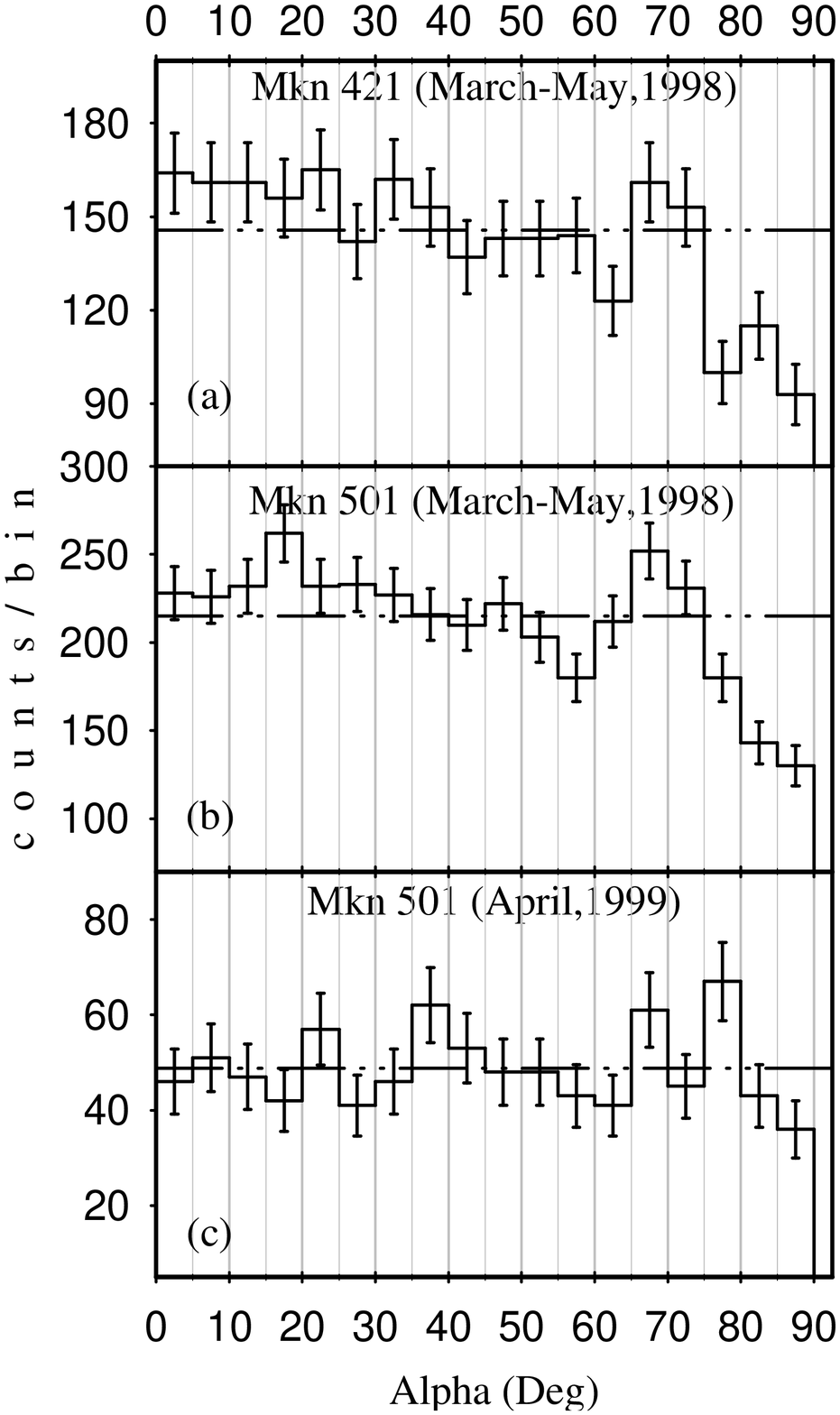,width=3.0in}}},%
{On source alpha-distributions for Mkn421 and Mkn501}]
Defining
$\alpha < 15^o$ as the $\gamma$-ray domain and $20^o \le \alpha \le
65^o$ as the control region, the event excess in the $\gamma$-ray domain
with respect to the average background value in the region 0 $\le \alpha \le 15^o$,
determined from the control region data, was estimated, both, on a
day-to-day basis and over the entire data-set. 
Figs. 1a and 1b represent the $\alpha$-distribution of the
image-parameter selected events for the two sources for the overall data.  In both the cases,
no significant excess of events is observed in the gamma-ray domain
($\alpha < 15^o$). The implied null
results lead to 3 $\sigma$ confidence level upper limits of
1.76 $\times 10^{-12}$
photons $cm^{-2}~s^{-1}$ for Mkn 421 and 1.4 $\times 10^{-12}$
photons $cm^{-2} s^{-1}$
for Mkn 501, above $E_\gamma >$ 2 TeV, during April-May, 1998,
after taking due account of the restricted field-of-view ($2.8^o
\times 1.2^o$) of the 4 $\times$ 9 pixel trigger generator used
in the 1998 observations.
%\end{figwindow}
%\subsection{1999 data set:}
Mkn 501 was observed for a total on-source observation time of $\sim$ 16.2h
over a period of 11 days spanning the period April 12 - 22, 1999.  The
same analysis procedure, as outlined above, was followed to search for a TeV
gamma-ray signal from the source during the period of our
observations.  The results are again summarized in the Table .
The $\alpha$ plot of the image parameter-selected events is
shown in Fig.1c; again no enhancement is evident in the $\gamma$-ray
event domain .
%\end{figwindow}
As seen from the
Table , the total number of events in the $\gamma$-domain ($\alpha \le
15^o$) is 144.00 $\pm$ 12.00 as compared to a background rate (RBL) of
146.33 $\pm$ 6.98, indicating the absence of a TeV signal from the source
during our epoch of observations. Based on these results a
3 $\sigma$ upper limit of 1.57 $\times 10^{-12}$ photons cm$^{-2}
s^{-1}$ above 2 TeV can be set on the source emission during
the 1999 epoch of observations.The analysis of the 1999 Mkn 421 data-set is currently on.
\end{figwindow}
\section{Discussion and Conclusions:}
Mkn 421 was first detected as a TeV gamma-ray source by the Whipple system
in 1991/92 (Punch et al, 1992) at a flux level of (1.59 $\pm$ 0.20)
$\times 10^{-11}$ photons cm$^{-2} s^{-1}$ above 500 GeV and subsequently
confirmed from observations carried out during 1993/1994 when its average
flux had dropped to $\sim 50\%$ of the 1992 level.  The most remarkable
observation concerning the TeV gamma-ray emission from Mkn 421 was the
detection of an intense burst between May 11-15, 1994 and a second burst
on June 7, 1994 by the Whipple system (Kerrick et al, 1995).
During the 1997
observing season, Mkn 421 was reported to be in a low state by a number of
imaging systems (Petry, 1997).  Although Mkn 421 is now considered to be a
firmly established TeV gamma-ray source, the nature of its short-term
variability and the exact shape of its energy spectrum in the VHE region
remain unclear due to mainly its relatively low intensity in the quiescent
state and the large systematic errors of spectral measurements with
Cerenkov telescopes.  The present observations indicate a continuation of
the low TeV activity state during the 1998 observation cycle.
%\end{figwindow}
Mkn 501 was first discovered as a VHE source by the Whipple system in 1995
at a level of 80 mCrab above 300 GeV (Quinn et al, 1996), and was
subsequently confirmed by the HEGRA system during observations carried out
in 1996 (Bradbury et al, 1997) above 1.5 TeV.  A large number of VHE
detections of Mkn 501 were reported at the 25th ICRC (Bhat,1997 ;Protheroe et al,
1997) including the first-ever near simultaneous detection of short-term flaring
activity by 5 imaging telescopes located across the globe.
The present observations indicate that the source
has relapsed into a low activity state during March - April, 1998
after the unusually active phase
witnessed in 1997 when it became the brightest object in VHE gamma-ray
sky.  As in the case of Mkn 421, the origin of short-term variability and
the prolonged phases of `low' activity remains unclear at present.

\vspace{1ex}
\begin{center}
{\Large\bf References}
\end{center}
Bhat, C.L. 1996, Proc. "Perspectives in High Energy Astronomy and
Astrophysics", Mumbai (India). University Press\\
Bhat, C.L. et al., 1997, Proc. "Towards a Major Atmospheric Cerenkov
Detector-V", Kruger, Park, South Africa, 196\\
Bhat, C.L. et al., 1994, NIM, A 340, 413\\
Bhat, C.L., 1997 , Proc. 25th ICRC, Durban, Rapporteur paper
Bradbury, S.M. et al., 1997, A $\&$ A, 320, L5\\
Koul, R. et al., 1997, Proc. "Towards a Major Atmospheric Cerenkov
Detector-V", Kruger Park, South Africa, 335\\
Kerrick, A.D. et al., 1995, ApJ, 438, L59\\
Petry, D. et al., 1997, Proc. 25th ICRC, Durban, 3, 241\\
Protheroe, R.J. et al, 1997, Proc. 25th ICRC (Invited, Rapporteur
and Highlight Papers), Durban (South Africa), 8, 317\\
Punch, M. et al., 1993, Nature, 160, 477\\
Quinn, J. et al., 1996, ApJ, 456, L83\\
Sapru, M.L. et al., 1997, Proc. "Towards a Major Atmospheric
Cerenkov Detector-V", Kruger Park, South Africa, 329\\
\end{document}